\documentclass[letterpaper, journal]{IEEEtran}

\usepackage{graphicx}
\usepackage{dcolumn}
\usepackage[table-column-type=Q]{siunitx}
\usepackage{adjustbox}

\usepackage{tabularx}
\usepackage{makecell}
\usepackage{array}

\usepackage{soul}
\usepackage{tikz}
\usepackage{amssymb}

\usepackage{cite}

\PassOptionsToPackage{hyphens}{url}\usepackage[hidelinks]{hyperref}

\ifCLASSINFOpdf
\else
\fi
\usepackage{amsmath}
\ifCLASSOPTIONcompsoc
 \usepackage[caption=false,font=normalsize,labelfont=sf,textfont=sf]{subfig}
\else
 \usepackage[caption=false,font=footnotesize]{subfig}
\fi
\usepackage[dvipsnames]{xcolor}

\usepackage{stfloats}

\usepackage[compact]{titlesec}

\usepackage{caption}
\usepackage[T1]{fontenc}

\newcommand*\sub[1]{_{\mathrm{#1}}}
\DeclareRobustCommand\_{\ifmmode\expandafter\sub\else\textunderscore\fi}

\newcolumntype{d}[1]{D..{#1}}

\definecolor{Gray}{gray}{0.85}
\newcolumntype{a}[1]{>{\columncolor{Gray}}Q{#1}}

\newcolumntype{L}{>{\centering\arraybackslash$}l<{$}}
\newcolumntype{R}{>{\centering\arraybackslash$}r<{$}}
\newcolumntype{C}{>{\centering\arraybackslash$}c<{$}}

\makeatletter
\def\mycol{}
\newcommand{\processeq}[1]{%
  \@processeq#1\relax
  \eqmakebox[lhs\mycol][r]{$\lhs$}%
  ${}={}$%
  \eqmakebox[rhs\mycol][l]{$\rhs$}%
}
\def\@processeq$#1=#2$\relax{\def\lhs{#1}\def\rhs{#2}}
\makeatother

\definecolor{colorpink}{HTML}{f3159e}
\definecolor{colorblue}{HTML}{11e7e7}
\definecolor{colorgray}{HTML}{afafaf}

\DeclareRobustCommand{\pinkline}{\raisebox{1.3pt}{\tikz{\draw[line cap=round, colorpink, dashdotted,  dash pattern=on 3pt off 3pt on .7pt off 3pt, line width=2pt](0,0) -- (12.6pt,0);}}}
\DeclareRobustCommand{\coloredline}[1]{\raisebox{1.3pt}{\tikz{\draw[line cap=round, #1, solid, line width=2pt](0,0) -- (12.6pt,0);}}}

\newcommand*\mystrut[1]{\vrule width0pt height0pt depth#1\relax}

\counterwithout{equation}{section}

\begin{document}

\title{\huge Influence of Cathode Boundary and Initial Electron Swarm Width on Electron Swarm Parameter Determination with the Pulsed Townsend Experiment}

\author{\vspace*{5pt}\IEEEauthorblockN{Mücahid Akbas\raisebox{2pt}{\small$^*$}}\vspace*{-5pt}%
\thanks{$^*$Part of this research was conducted while the author was affiliated with the High Voltage Laboratory, ETH Zürich, Switzerland (e-mail: makbas@ethz.ch).}

\IEEEauthorblockA{%
}%
}%

\maketitle

\begin{abstract}

The Pulsed Townsend experiment enables the extraction of relevant electron swarm transport properties in different gases such as the electron drift velocity $W$ (or equivalently the mobility $\mu$), the longitudinal diffusion coefficient $D\sub{L}$, and the effective ionization rate $R\sub{net}$ (or equivalently the effective ionization coefficient $\alpha\sub{eff}$). Existing analysis techniques lack an accurate representation of the experimental initial and boundary conditions. This work provides an improved evaluation approach by appropriately considering both initial and boundary conditions in order to extract more accurate swarm parameters from measurement data. Simulative and experimental measurement results verify an increased evaluation accuracy. Furthermore, the longitudinal diffusion coefficient $D\sub{L}$ can now be accurately extracted from Pulsed Townsend measurements, which was previously not possible with existing evaluation approaches. The developed curve fitting code is made publicly available.

\end{abstract}


\section{Introduction}

\begin{figure*}[b!]
	\centering
	\includegraphics[width=0.7\linewidth]{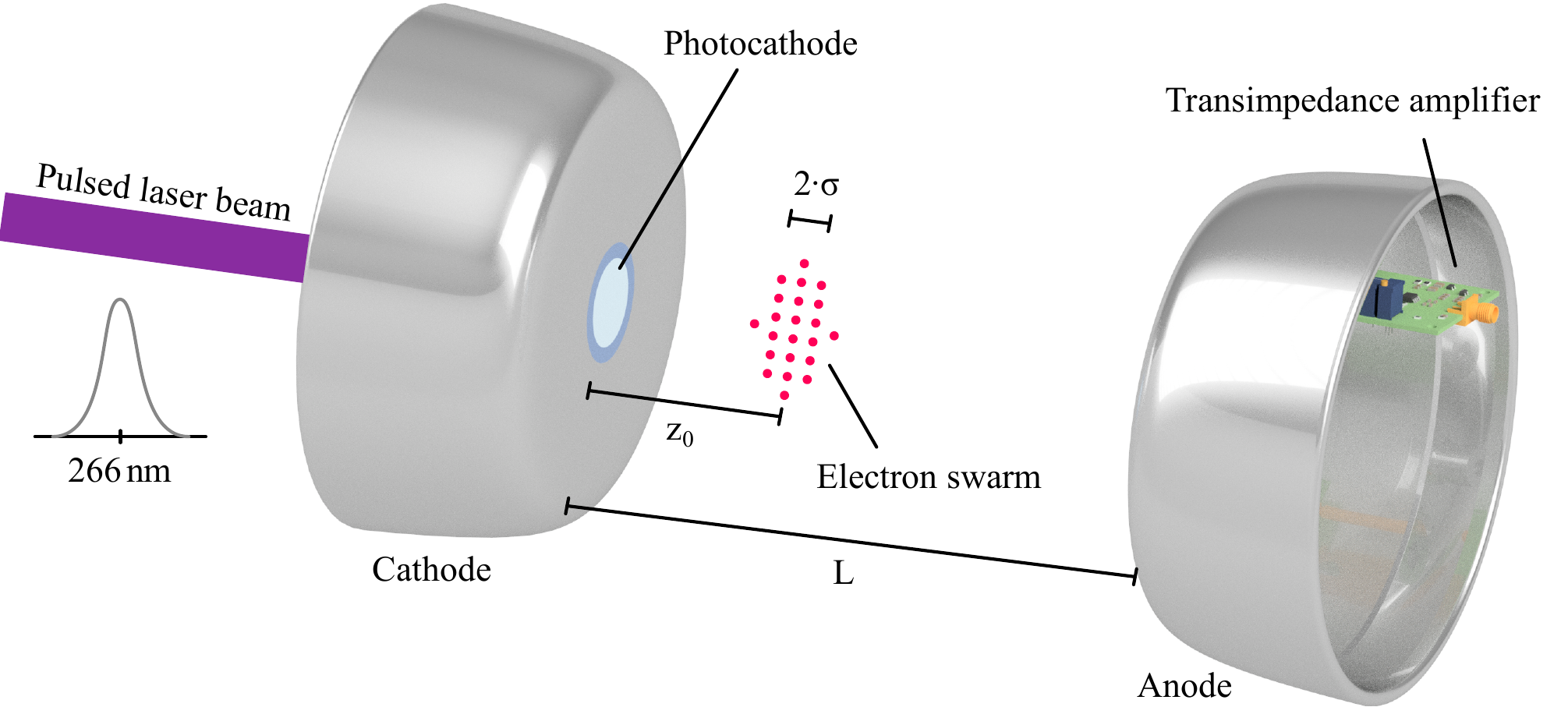}
	\caption{ Electrode arrangement and initial electron swarm, which is assumed to be distributed as a Gaussian pulse with $\sigma$ denoting the Gaussian RMS width in longitudinal direction, and $z\sub{0}$ the initial (peak) position of released electrons. The electrodes have a total diameter of $d\sub{total} = \SI{165}{mm}$ with the flat surface being around $d\sub{flat} = \SI{114}{mm}$ wide, and a gap spacing of around $L = 10\dots\SI{35}{mm}$ \cite{Haefliger_thesis}. The photocathode has a PdZn nanocrystalline coating \cite{Photocathode} and is $\SI{25}{mm}$ wide. A custom-made electrode mounted transimpedance amplifier is included for illustrative purposes. }
	\label{fig:electrode_arrangement}
\end{figure*}

\noindent Electron and ion swarm parameters in gaseous media constitute key input data for gas discharge models, plasma physics and devices \cite{Petrovic_2009, Napartovich_2011}, and atmospheric sciences \cite{Petrovic_2017}. Different experimental arrangements such as the Steady-State Townsend (SST), the Time-of-Flight (ToF) (scanning drift tube \cite{Korolov_2016, Vass_2021}) and the Pulsed Townsend (PT) \cite{Hao_2024, Dahl_2012, Sasic_2010} experiment exist to determine some of these transport properties. The Pulsed Townsend experiment in particular enables determining swarm parameters for both electrons and ions given suitable models and fitting techniques. The relevant transport properties comprise the (bulk) drift velocity $W^{\mathrm{b}}$, (bulk) longitudinal diffusion coefficient $D\sub{L}^{\mathrm{b}}$ and the effective ionization rate $R\sub{net}$ (or equivalently the effective ionization coefficient $\alpha\sub{eff}$).

In the Pulsed Townsend experiment, the displacement currents of pulsed electron avalanches in the (lower) Townsend regime are measured and analyzed. A pulsed ultra-violet (UV) laser is used to release initial electrons from a photocathode, see \textbf{Fig.~\ref{fig:electrode_arrangement}}. These are subsequently accelerated by an applied constant (DC) homogeneous electric field, and thus, drift towards the anode. The resulting capacitive current is measured using a transimpedance amplifier and captured with an oscilloscope. Using an analytical expression for this current signal and curve fitting, the underlying swarm parameters are extracted.

In literature, existing analytical models do not impose an absorbing cathode boundary condition and are, as such, considering either the case of an assumed infinite (drift) domain \cite{Chachereau_2016} or half-infinite domain with an absorbing anode boundary \cite{Casey_2021}. Thus, effects such as electron back-diffusion to the cathode or a potential asymmetry of the early-time electron distribution are not represented by these models. Furthermore, experimental limitations \cite{PT-Haefliger} such as a finite laser pulse width or a finite measurement bandwidth that result in a broadened electron current waveform have not been appropriately included in the expressions and fitting routines so far. The effect is especially critical for $D\sub{L}^{\mathrm{b}}$, because diffusion is primarily inferred from the swarm broadening that affects the tail of a measured current waveform.

The present work provides a new evaluation approach in order to extract more accurate swarm parameter data from Pulsed Townsend current waveforms. This approach involves a new finite domain analytical expression and accounts for both a finite laser pulse width and a limited measurement bandwidth. Numerical and experimental validation highlight the resulting impact on swarm parameter determination. Finally, the developed fitting code is made publicly available to increase transparency into swarm parameter measurements and enable more accurate measurements across different research groups.

\section{The Pulsed Townsend Experiment}

\subsection{Theory}

\noindent In order to obtain electron swarm transport data from measurements of a swarm experiment (e.g., the Pulsed Townsend experiment) a suitable physical model is necessary. A Boltzmann equation analysis \cite{Ron_Book, Tagashira} provides an accurate description of the microscopic electron dynamics within an electron avalanche process. However, such an analysis is complicated and time consuming, and thus, practically not directly applicable to extract swarm parameters from swarm experiments. Traditionally, fluid equations (e.g., the diffusion equation) are utilized \cite{Chen_Book, Casey_2021, Ron_Book} to make swarm parameters more accessible from measurements. The resulting (one-dimensional) electron number density continuity equation can generally be written as \cite{Tagashira}
\setcounter{equation}{0}%
\begin{equation}\label{eq:n_continuity}
	\dfrac{\partial n\sub{e}}{\partial t} + W^{\mathrm{b}}\cdot \dfrac{\partial n\sub{e}}{\partial z} - D\sub{L}^{\mathrm{b}}\cdot \dfrac{\partial^2 n\sub{e}}{\partial z^2} + Q\sub{L}^{\mathrm{b}}\cdot \dfrac{\partial^3 n\sub{e}}{\partial z^3} \mp \dots = R\sub{net}\cdot n\sub{e} \,\text{,}
\end{equation}
where $n\sub{e}$ denotes the electron number density, $W^{\mathrm{b}}$ the bulk drift velocity, $D\sub{L}^{\mathrm{b}}$ the bulk longitudinal diffusion coefficient, $Q\sub{L}^{\mathrm{b}}$ the bulk longitudinal skewness coefficient (3$^{\text{rd}}$ order term), and $R\sub{net}$ the effective ionization rate (including ionization and attachment rate coefficients). This partial differential equation may be solved under given initial (e.g., $n\sub{e}(z, t = 0) = n\sub{e,0}\cdot \delta(z)$) and boundary conditions (e.g., cathode at $z = -\infty$ and anode at $z = L$) to obtain an expression for the density profile $n\sub{e}(z, t)$ over time and space.

Key notable physical assumptions here include hydrodynamic conditions (i.e., the weak-gradient limit), no electron-electron interactions (thus neglecting potential space-charge effects), and that the coefficients be time independent \cite{Tagashira}.

In order to obtain an expression for the externally measurable electron current $I\sub{e}(t)$ (observable), Fick's law needs to be considered \cite{Casey_2021}
\setcounter{equation}{1}%
\begin{equation}
	\Gamma(z, t) = W^{\mathrm{f}}\cdot n\sub{e} - D\sub{L}^{\mathrm{f}}\cdot\dfrac{\partial n\sub{e}}{\partial z} + Q\sub{L}^{\mathrm{f}}\cdot\dfrac{\partial^2 n\sub{e}}{\partial z^2} \mp \dots \,\text{,}
\end{equation}
where $\Gamma(z, t)$ denotes the particle flux, and quantities with superscript 'f' represent flux swarm parameters \cite{Sakai, Ron_Book}. An expression for a measurable displacement current is, thus, found as \cite{Casey_2021}
\setcounter{equation}{2}%
\begin{equation}
	I\sub{e}(t) = \dfrac{q}{L} \cdot \int_{z = 0}^{z = L} \Gamma(z, t) \;\mathrm{d}z \,\text{,}
\end{equation}
where $q$ denotes the electronic charge, and $L$ the gap distance.

\subsection{Simulation}\label{sec:fluid-code}

\begin{figure}[b!]
	\centering
	\includegraphics[width=1\linewidth, trim={0 0 0 2.3cm}, clip]{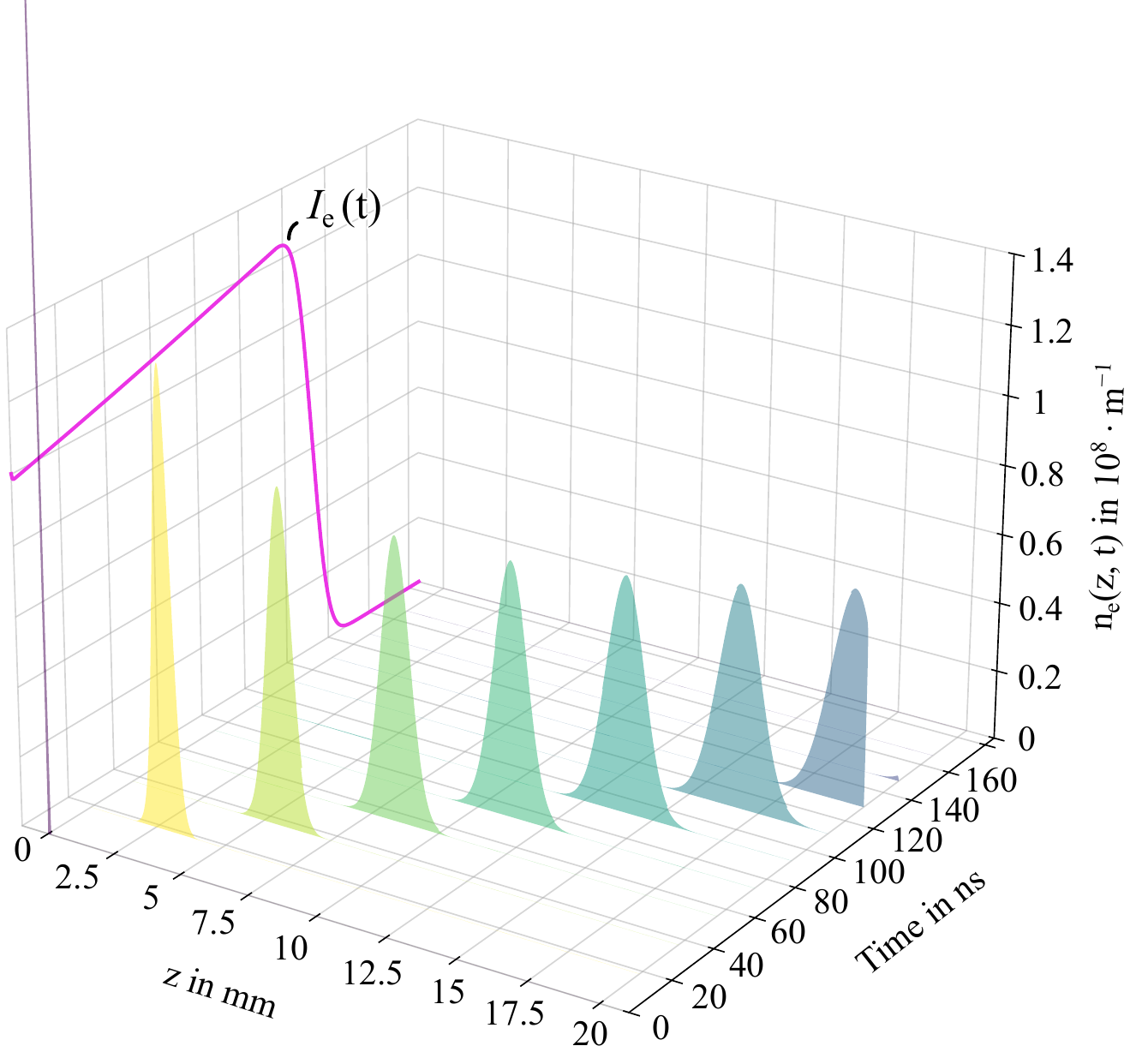}
	\caption{ Example electron number density profile over space and time within an electrode gap ($L = \SI{20}{mm}$) for net ionizing conditions. The resulting externally measurable current $I\sub{e}(t)$ (in arbitrary units) is also illustrated. }
	\label{fig:n_evolution}
\end{figure}

\noindent This section shortly introduces a fluid code that is employed in this work in order to obtain accurate simulated electron current waveforms for a Pulsed Townsend experiment setting. Essentially, the continuity equation \eqref{eq:n_continuity} is numerically solved given suitable initial and boundary conditions, and optionally including an arbitrary number of higher order terms (e.g., skewness $Q\sub{L}^{\mathrm{b}}$). The spatial domain is discretized in the z-axis (e.g., in $1\dots\SI{10}{\micro\meter}$ steps) and the spatial derivatives of $n\sub{e}(z, t)$ are expressed by finite differences. The resulting ordinary differential equation in $t$ is numerically solved using an implicit Runge-Kutta solver, as the considered differential equation is very stiff even for moderate swarm parameter values. \textbf{Fig.~\ref{fig:n_evolution}} shows an example simulation for net ionizing conditions and a gap distance of $L = \SI{20}{mm}$.

\section{State-of-the-art Evaluation Approach}

\setcounter{equation}{6}%
\begin{figure*}[t!]
	\begin{multline}\label{eq:casey-eq}
		I\sub{e}(t) = \dfrac{n\sub{e,0} \cdot q \cdot W^{\mathrm{f}}}{2\cdot L} \cdot \exp\bigl( R\sub{net}\cdot (t - t\sub{0}) \bigr) \cdot \\
		\Biggl\{ \Biggl[ 1 - \mathrm{erf}\Biggl( \dfrac{(t - t\sub{0}) - T\sub{e}}{\sqrt{2 \cdot \tau\sub{D} \cdot (t - t\sub{0})}} \Biggr) \Biggr]  +  \exp\Biggl( \dfrac{2\cdot T\sub{e}}{\tau\sub{D}} \Biggr) \cdot \Biggl[ \mathrm{erf}\Biggl( \dfrac{(t - t\sub{0}) + T\sub{e}}{\sqrt{2\cdot \tau\sub{D} \cdot (t - t\sub{0})}} \Biggr) - 1 \Biggr] \Biggr\}\raisebox{-8pt}{.}
	\end{multline}
	\hrulefill
	\vspace*{-0.3cm}
\end{figure*}

\setcounter{equation}{9}%
\begin{figure*}[b!]
	\vspace*{-0.3cm}
	\hrulefill
	\begin{align}\label{eq:number-density}
		n\sub{e}(z, t) = &\dfrac{n\sub{e,0} \cdot \exp\Biggl( R\sub{net} \cdot (t - t\sub{0}) + \frac{W^{\mathrm{b}}}{2\cdot D\sub{L}^{\mathrm{b}}} \cdot (z - z\sub{0} - \frac{1}{2}\cdot W^{\mathrm{b}}\cdot (t - t\sub{0})) \Biggr)}{\sqrt{4\pi \cdot D\sub{L}^{\mathrm{b}}\cdot (t-t\sub{0})}} \cdot \nonumber\\
		&\sum_{j = -\infty}^{j = +\infty} \Biggl[ \exp\Biggl( -\dfrac{(z - z\sub{0} - 2j\cdot L)^2}{4\cdot D\sub{L}^{\mathrm{b}}\cdot (t - t\sub{0})} \Biggr) - \exp\Biggl( -\dfrac{(z + z\sub{0} - 2j\cdot L)^2}{4\cdot D\sub{L}^{\mathrm{b}} \cdot (t - t\sub{0})} \Biggr)  \Biggr]\raisebox{-8pt}{,}
	\end{align}
	\\[-10pt]
	\setcounter{equation}{10}%
	\begin{align}\label{eq:back-diffusion}
		I\sub{e}(t) = &\dfrac{n\sub{e,0} \cdot q \cdot W^{\mathrm{f}}}{2 \cdot L} \cdot \sum_{j = -\infty}^{+\infty}\Biggl\{ ~\exp\Bigl(R\sub{net}\cdot (t - t\sub{0}) + 2 \cdot j \cdot \dfrac{T\sub{e}}{\tau\sub{D}}\Bigr) ~ \cdot \nonumber\\ 
		\Biggl[ ~ \mathrm{erf}&\Biggl( \dfrac{1}{\sqrt{2\cdot\tau\sub{D}}}\cdot \biggl(\dfrac{T\sub{e}\cdot(1 - 2\cdot j)}{\sqrt{t - t\sub{0}}} - \sqrt{t - t\sub{0}} - \dfrac{T\sub{z}}{\sqrt{t - t\sub{0}}}\biggr) \Biggr) - \mathrm{erf}\Biggl( \dfrac{-1}{\sqrt{2\cdot\tau\sub{D}}}\cdot \biggl(\sqrt{t - t\sub{0}} + \dfrac{T\sub{z} + 2\cdot j\cdot T\sub{e}}{\sqrt{t - t\sub{0}}}\biggr) \Biggr) ~ \Biggr] \nonumber\\
		- \exp &\Bigl( R\sub{net}\cdot (t - t\sub{0}) + 2\cdot j\cdot \dfrac{T\sub{e}}{\tau\sub{D}} - 2\cdot\dfrac{T\sub{z}}{\tau\sub{D}} \Bigr) ~ \cdot \nonumber\\
		\Biggl[ ~ \mathrm{erf}&\Biggl( \dfrac{1}{\sqrt{2\cdot\tau\sub{D}}}\cdot \biggl(\dfrac{T\sub{e}\cdot(1 - 2\cdot j)}{\sqrt{t - t\sub{0}}} - \sqrt{t - t\sub{0}} + \dfrac{T\sub{z}}{\sqrt{t - t\sub{0}}}\biggr) \Biggr) - \mathrm{erf}\Biggl( \dfrac{-1}{\sqrt{2\cdot\tau\sub{D}}}\cdot \biggl(\sqrt{t - t\sub{0}} + \dfrac{-T\sub{z} + 2\cdot j\cdot T\sub{e}}{\sqrt{t - t\sub{0}}}\biggr) \Biggr) ~ \Biggr] \Biggr\}\raisebox{-8pt}{.}
	\end{align}
\end{figure*}

\begin{figure*}[t!]
	\centering
	\includegraphics[width=1.\linewidth]{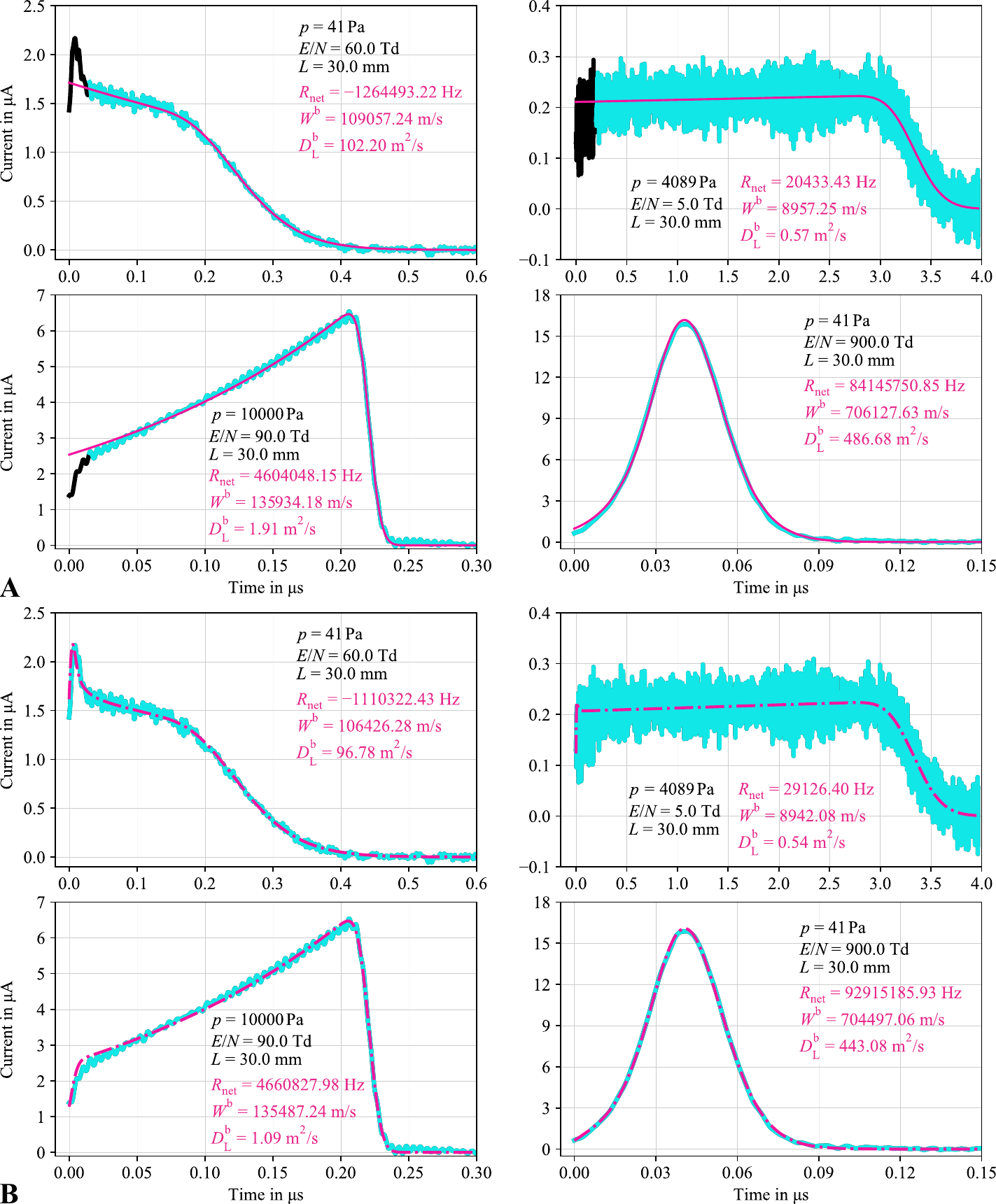}
	\caption{ Example experimental curve fitting results for CO$\sub{2}$ gas under different conditions using \textbf{(A)} the state-of-the-art approach, and \textbf{(B)} the proposed method. The black curve {\coloredline{black}} represents the raw experimental waveform, the blue curve {\coloredline{colorblue}} is used for fitting and results from the raw waveform by cutting and downsampling, the pink curve {\pinkline} shows the obtained fit. (Please note that some of the shown raw experimental waveforms of the author have been made openly available and can be found at \cite{Zenodo-PT-Paper-HVL}, and some related swarm data over LXCat \cite{CO2_LXCat}). }
	\label{fig:example-fitting}
\end{figure*}

\noindent In the past, curve fitting of Pulsed Townsend electron current waveforms was performed using the following analytical expression \cite{Alise_thesis, Chachereau_2016, Pachin_2019, Haefliger_thesis}:
\setcounter{equation}{3}%
\begin{align}
	&I\sub{e}(t) = \dfrac{n\sub{e,0} \cdot q \cdot W}{2\cdot L} \cdot \mathrm{exp}\Bigl(R\sub{net}\cdot (t - t\sub{0})\Bigr) \cdot \nonumber\\[4pt]
	&\Biggl[ \underbrace{\mystrut{3.5ex} \mathrm{erf}\Biggl( \dfrac{t - t\sub{0}}{\sqrt{2\cdot \tau\sub{D} \cdot (t - t\sub{0})}} \Biggr)}_{\approx 1} -\, \mathrm{erf}\Biggl( \dfrac{(t - t\sub{0}) - T\sub{e}}{\sqrt{2\cdot \tau\sub{D}\cdot (t - t\sub{0})}} \Biggr)  \Biggr]\raisebox{-8pt}{,}
\end{align}
where $T\sub{e} = \dfrac{L}{W^{\mathrm{b}}}$ denotes the electron arrival time at the anode, $\tau\sub{D} = \dfrac{2\cdot D\sub{L}^{\mathrm{b}}}{{W^{\mathrm{b}}}^2}$ the characteristic time scale for electron diffusion, and $t\sub{0}$ an optional time shift (see section \ref{sec:fitting}).
The underlying model is given by a solution to the drift-diffusion equation (i.e., equation~\eqref{eq:n_continuity} with terms up to second order) for an infinite domain with both the cathode and anode positioned at infinite distance ($z = -\infty$ and $z = \infty$, respectively) from the (initial) electron swarm, where the initial electron swarm is (for simplicity) further assumed to be released as a Dirac pulse (in space and time) of $n\sub{e,0}$ particles
\setcounter{equation}{4}%
\begin{equation}\label{eq:initial}
	n\sub{e}(z, t = t\sub{0}) = n\sub{e,0}\cdot\delta(z) \,\text{.}
\end{equation}
The resulting electron number density profile $n\sub{e}(z, t)$ can be denoted as \cite{Ron_Book}:
\setcounter{equation}{5}%
\begin{align}
    n\sub{e}(z, t) = &\dfrac{n\sub{e,0} \cdot \exp\bigl( R\sub{net} \cdot (t - t\sub{0}) \bigr) }{ \sqrt{ 4\pi\cdot D\sub{L}^{\mathrm{b}} \cdot (t - t\sub{0}) } } \cdot \nonumber\\
    &\exp\Biggl( -\dfrac{(z - W^{\mathrm{b}}\cdot (t - t\sub{0}))^2}{4\cdot D\sub{L}^{\mathrm{b}} \cdot (t - t\sub{0})} \Biggr)\raisebox{-8pt}{.}
\end{align}

Recently, Casey et al. \cite{Casey_2021} have presented an electron current expression for the half-infinite domain case, where the cathode is placed at $z = -\infty$ and the anode at $z = L$, see equation \eqref{eq:casey-eq}. A more accurate swarm parameter extraction is enabled by using this expression for curve fitting \cite{Hanut_PT, Marnik_PT}. Example curve fitting results are illustrated in \textbf{Fig.~\ref{fig:example-fitting}A}. Furthermore, Casey et al. \cite{Casey_2021} provided a detailed discussion on the role and differentiation of flux and bulk swarm parameters, which is not repeated here in the interest of brevity.

\section{Improved Evaluation Approach}

\noindent In order to further improve the accuracy of extracted swarm parameters, this section first presents an analytical solution for the electron current $I\sub{e}(t)$ considering a finite domain with both an absorbing cathode and anode boundary at $z = 0$ and $z = L$, respectively. Additionally, a Gaussian distribution (in time) is considered and numerically included as an initial condition in order to more accurately represent observed experimental conditions. 

A solution to the drift-diffusion equation for an initial condition as $n\sub{e}(z, t = t\sub{0}) = n\sub{e,0}\cdot \delta(z - z\sub{0})$ and for absorbing boundaries is provided in equation~\eqref{eq:number-density} \cite{Robson}. The corresponding electron current signal can be determined by (suitable) integration over $z$ as (considering terms only up to second order)
\setcounter{equation}{7}%
\begin{equation}\label{eq:current-integral}
    I\sub{e}(t) = \dfrac{q\cdot W^{\mathrm{f}}}{L} \cdot \int_{z = 0}^{z = L} n\sub{e}(t, z) \;\mathrm{d}z \,\text{.}
\end{equation}
By appropriately collecting terms in equation~\eqref{eq:number-density} and by utilizing the identity
\setcounter{equation}{8}%
\begin{align}
    &\int \exp\Bigl( a + b \cdot z - c \cdot z^2 \Bigr) \;\mathrm{d}z = \nonumber\\
    &\dfrac{\sqrt{\pi}\cdot \exp\Biggl(a + \dfrac{b^2}{4\cdot c}\Biggr) \cdot \mathrm{erf}\Biggl(\dfrac{2\cdot c \cdot z - b}{2\cdot \sqrt{c}}\Biggr)}{2\cdot \sqrt{c}},
\end{align}
equations \eqref{eq:current-integral} and \eqref{eq:number-density} result in equation~\eqref{eq:back-diffusion} for the electron current waveform, where $T\sub{e} = \dfrac{L}{W^{\mathrm{b}}}$ denotes the electron arrival time at the anode, $0 < T\sub{z} = \dfrac{z\sub{0}}{W^{\mathrm{b}}} < T\sub{e}$ a time quantity that is related to the initial spatial starting position of the electrons, $\tau\sub{D} = \dfrac{2\cdot D\sub{L}^{\mathrm{b}}}{{W^{\mathrm{b}}}^2}$ the characteristic time scale for electron diffusion.
Due to the fast convergence of the infinite series in equation~\eqref{eq:back-diffusion}, practically only a few terms (e.g., with $|j| \leq 2$) are required to obtain a sufficiently accurate result.

\textbf{Fig.~\ref{fig:electrode_arrangement}} indicates an initial (finite) distribution of the initial electron swarm. This initial spread is also visible in the resulting experimental current waveforms in \textbf{Fig.~\ref{fig:example-fitting}A}. Generally, there are two contributions to this spread $\sigma$: an initial spatial spread $\sigma\sub{z}$ and an initial temporal spread $\sigma\sub{t}$. For the experimental setup considered, it is found that (as discussed in section \ref{sec:improved-swarm-params}) the temporal contribution is dominating (under common measurement conditions, i.e., not too low fields of around $E/N = \SI{1}{Td}$ or lower).

The initial temporal pulse width $\sigma\sub{t}$ mainly results from a finite laser pulse width (typically in the order of a few $\SI{}{ns}$ \cite{Vass_2021, PT-Haefliger}) and limitations in the experimental measurement bandwidth. A finite laser pulse width physically contributes to $\sigma\sub{t}$, whereas a limited signal bandwidth only apparently contributes to $\sigma\sub{t}$. A further differentiation between the two contributions is avoided for simplicity, as an analysis with an 'apparent' or total $\sigma\sub{t}$ provides the same outcome as an analysis that considers the contributions separately. Please note that higher order effects such as potential non-idealities in the transfer-function of the measurement loop (i.e., transimpedance amplifier, cabling, electrode arrangement, etc.) are thus neglected.

For simplicity, the pulse shape is assumed and modelled by a Gaussian function
\setcounter{equation}{11}%
\begin{equation}\label{eq:gaussian}
    f(t) = \dfrac{1}{\sigma\sub{t} \cdot \sqrt{2\pi}} \cdot \exp\Biggl( -\dfrac{t^2}{2\cdot {\sigma\sub{t}}^2} \Biggr)\raisebox{-8pt}{,}
\end{equation}
where $\sigma\sub{t}$ represents the RMS width and is related to the full width at half maximum (FWHM) by $\mathrm{FWHM} = \sqrt{8\cdot\ln(2)}\cdot \sigma\sub{t} \approx 2.35 \cdot \sigma\sub{t}$. This initial condition is in the simplest case incorporated via numerical convolution with the analytical electron current waveform $I\sub{e}(t)$
\setcounter{equation}{12}%
\begin{equation}\label{eq:convolution}
	I\sub{e,\,\sigma\sub{t}}(t) = f(t) \;*\; I\sub{e}(t) = \sum\sub{m = -\infty}^{\infty} f[m] \cdot I\sub{e}[k - m] \,\text{,}
\end{equation}
where the square brackets denote an indexing operation on discrete sequences.

\setcounter{equation}{13}%
\begin{figure*}[b!]
	\vspace*{-0.3cm}
	\hrulefill
	\begin{align}\label{eq:cost-function}
		J\sub{cost}(C, R\sub{net}, T\sub{e}, &\tau\sub{D}, T\sub{z}, t\sub{0}, \vec{t}, \vec{I\sub{e}}) = \nonumber\\[3pt]
		\Biggl|\Biggl| f(t) * \Biggl\{ C \cdot \sum_{j = -\infty}^{j = +\infty} &\exp\Biggl[ R\sub{net} \cdot (\vec{t} - t\sub{0}) + 2\cdot j\cdot \dfrac{T\sub{e}}{\tau\sub{D}} + \ln\Biggl( \mathrm{erf}\Biggl( \dfrac{1}{\sqrt{2\cdot\tau\sub{D}}}\cdot \biggl(\dfrac{T\sub{e}\cdot(1 - 2\cdot j)}{\sqrt{\vec{t} - t\sub{0}}} - \sqrt{\vec{t} - t\sub{0}} - \dfrac{T\sub{z}}{\sqrt{\vec{t} - t\sub{0}}}\biggr) \Biggr) \nonumber\\
		&\qquad\qquad\qquad\qquad\qquad\qquad\qquad\qquad\qquad- \mathrm{erf}\Biggl( \dfrac{-1}{\sqrt{2\cdot\tau\sub{D}}}\cdot \biggl(\sqrt{\vec{t} - t\sub{0}} + \dfrac{T\sub{z} + 2\cdot j\cdot T\sub{e}}{\sqrt{\vec{t} - t\sub{0}}}\biggr) \Biggr) \Biggr) \Bigg] \nonumber\\
		- ~ &\exp \Biggl[ R\sub{net} \cdot (\vec{t} - t\sub{0}) + 2\cdot j\cdot \dfrac{T\sub{e}}{\tau\sub{D}} - 2\cdot\dfrac{T\sub{z}}{\tau\sub{D}} +  \ln\Biggl( \mathrm{erf}\Biggl( \dfrac{1}{\sqrt{2\cdot\tau\sub{D}}}\cdot \biggl(\dfrac{T\sub{e}\cdot(1 - 2\cdot j)}{\sqrt{\vec{t} - t\sub{0}}} - \sqrt{\vec{t} - t\sub{0}} + \dfrac{T\sub{z}}{\sqrt{\vec{t} - t\sub{0}}}\biggr) \Biggr) \nonumber\\
		&\qquad\qquad\qquad\qquad\qquad\qquad\qquad\qquad\qquad- \mathrm{erf}\Biggl( \dfrac{-1}{\sqrt{2\cdot\tau\sub{D}}}\cdot \biggl(\sqrt{\vec{t} - t\sub{0}} + \dfrac{-T\sub{z} + 2\cdot j\cdot T\sub{e}}{\sqrt{\vec{t} - t\sub{0}}}\biggr) \Biggr) \Biggr) \Bigg] \Biggr\} \nonumber\\
		- &\dfrac{1}{I\sub{e,max}} \cdot \vec{I\sub{e}} ~ \Biggr|\Biggr|\raisebox{-8pt}{,} \\[3pt]
		\text{where $||\cdot||$ denotes} &\text{ either the $||\cdot||\sub{1}$ or $||\cdot||\sub{2}$ norm, and } f(t) \text{ the Gaussian function in equation~\eqref{eq:gaussian}.} \nonumber\\[-22pt]\nonumber
	\end{align}
\end{figure*}

In \textbf{Fig.~\ref{fig:example-fitting}B}, example curve fitting results are illustrated using equation \eqref{eq:back-diffusion} and \eqref{eq:convolution}. The proposed model reproduces the early-time decay and arrival-time broadening more precisely, and thus, more accurate swarm parameters are obtained, as further discussed in section \ref{sec:improved-swarm-params}. Furthermore, the resulting characteristic feature from electron back-diffusion to the cathode can be reproduced by the proposed approach (equations~\eqref{eq:back-diffusion} and \eqref{eq:convolution}). This feature is most pronounced at lower pressures, and thus, larger longitudinal diffusion coefficients, and manifests itself in an initial decay in the current waveform as electrons are being absorbed at the cathode.

\section{Implementation}\label{sec:fitting}

\noindent The investigated curve fitting approaches have been implemented in Python utilizing the scientific libraries \textit{NumPy} \cite{numpy} and \textit{SciPy} \cite{scipy}. Reliable fitting results are achieved by using a global optimization method such as, e.g., \textit{differential\_evolution($\dots$)} and an appropriate cost function formulation using equation~\eqref{eq:back-diffusion} (and \eqref{eq:convolution})\footnote{See also a version of the previously made available code at \url{https://gitlab.com/ethz\_hvl/APTX}, last accessed on 31.01.26.}. A suitable example for an objective function is given in equation~\eqref{eq:cost-function}, where $C$ denotes an arbitrary constant of multiplication, $R\sub{net}$ the effective ionization rate, $T\sub{e} = \dfrac{L}{W^{\mathrm{b}}}$ the arrival time of electrons at the anode, $\tau\sub{D} = \dfrac{2\cdot D\sub{L}^{\mathrm{b}}}{{W^{\mathrm{b}}}^2}$ a characteristic time scale for longitudinal diffusion, $T\sub{z} = \dfrac{z\sub{0}}{W^{\mathrm{b}}}$ a time quantity related to the initial electron swarm position, $t\sub{0}$ an optional time offset to, e.g., match the time axes between the physical model and an experimental waveform, $\vec{t}$ the time vector for an electron current waveform, and $\vec{I\sub{e}}$ the corresponding waveform vector.

In order to ensure numerical stability and good fitting performance, parts of equation~\eqref{eq:back-diffusion} are expressed in log-space. Otherwise, the exponential terms $\mathrm{exp}\Bigl(2\cdot j\cdot \frac{T\sub{e}}{\tau\sub{D}}\Bigr)$ and $\mathrm{exp}\Bigl(-2\cdot\frac{T\sub{z}}{\tau\sub{D}}\Bigr)$ would become unstable for large ratios of $\frac{T\sub{e}}{\tau\sub{D}}$ or $\frac{T\sub{z}}{\tau\sub{D}}$ (practically for small diffusion coefficient values $D\sub{L}^{\mathrm{b}}$). Furthermore, for cases with very large $D\sub{L}^{\mathrm{b}}$ (e.g., observed at very low gas densities), the error function terms in equation~\eqref{eq:back-diffusion} could result in numerical 'jitter' or noise as the computations would be in the range of and below common floating point precision (e.g, double-precision floating point numbers with 64-bit size). Using higher precision floating point numbers in such cases would avoid potential numerical noise when fitting. Lastly, stating the electron current or cost function in terms of time domain parameters instead of directly using the electron swarm parameters enables choosing the optimization boundaries based on the time scale, i.e., the length of an individual waveform. Thus, the boundaries are gas and pressure independent and the curve fitting algorithm universally applicable, similarly to the implementation of the state-of-art approach used in \cite{PT-paper-HVL}. 

The implementation is made publicly available over GitHub (at \textit{https://github.com/mueakbas/pt-curve-fit}) in order to enable different research groups to obtain more accurate swarm parameters from Pulsed Townsend measurements and provide users of swarm transport data a better perception of the employed extraction tools.

\section{Swarm Parameter Accuracy}\label{sec:improved-swarm-params}

\begin{figure*}[t!]
	\centering
	\includegraphics[width=1\linewidth]{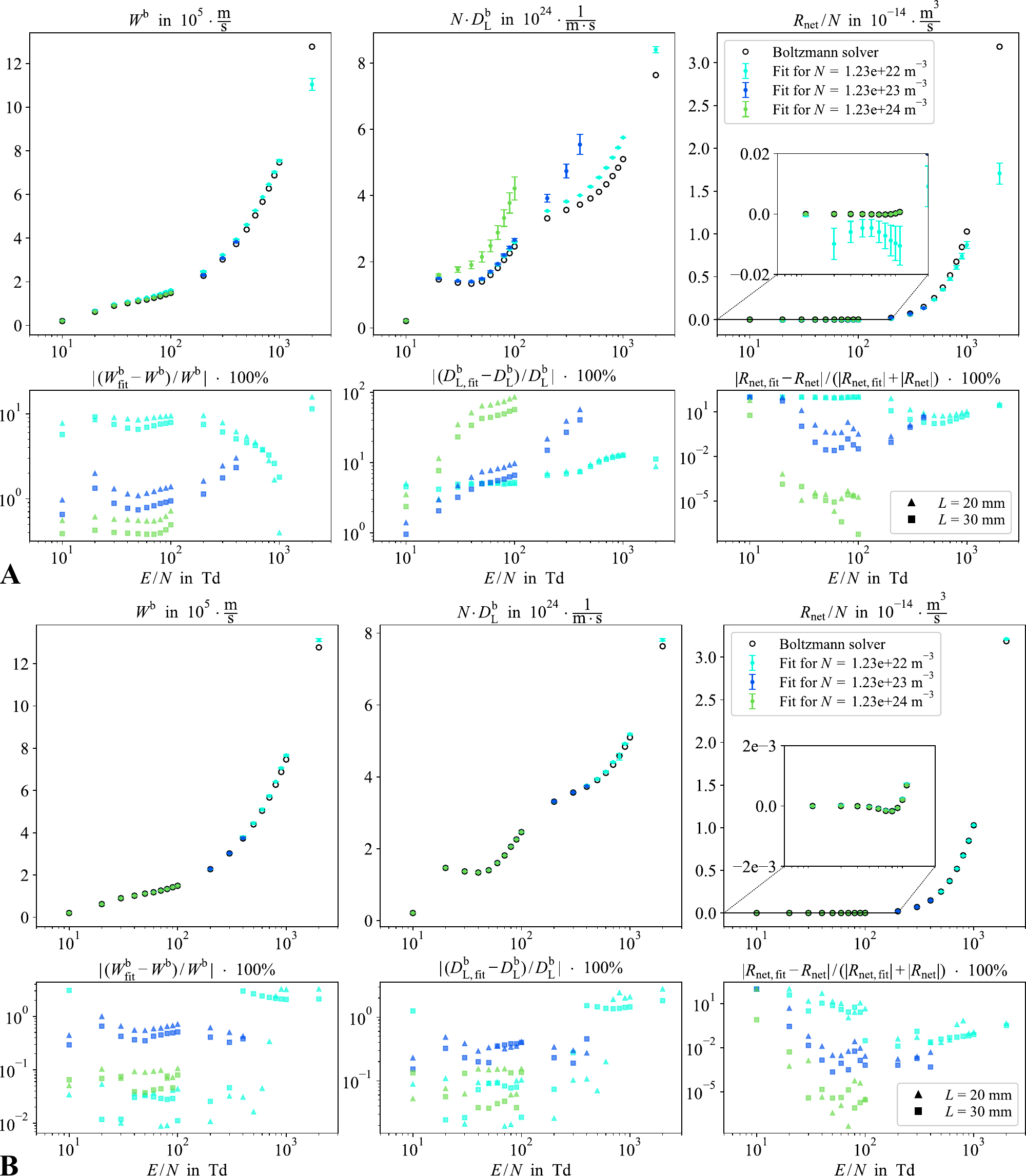}
	\caption{ Electron swarm parameters for simulated waveforms of CO$_{2}$ gas that are evaluated with \textbf{(A)} the existing method in equation~\eqref{eq:casey-eq}, and \textbf{(B)} the proposed method in equation~\eqref{eq:back-diffusion} and \eqref{eq:convolution}. The initial pulse width is chosen as $\sigma\sub{t} = \SI{4.5}{ns}$ and the initial swarm position around $z\sub{0} = \SI{100}{\micro\meter}$. The underlying swarm parameter values are calculated with MultiBolt \cite{Stephens_2018, Flynn_2022} using Biagi's cross section set for CO$\sub{2}$ \cite{Biagi_CO2_LXCat}. Electron current waveforms are simulated (see section \ref{sec:fluid-code}) for the three pressures $p = \SI{50}{Pa}$ ($N \approx \SI{1.23e22}{\per\meter\cubed}$), $p = \SI{500}{Pa}$ ($N \approx \SI{1.23e23}{\per\meter\cubed}$) and $p = \SI{5}{kPa}$ ($N \approx \SI{1.23e24}{\per\meter\cubed}$) over a field range of $E/N = 10\dots\SI{2000}{Td}$. The top row shows the evaluated swarm parameters and the underlying Boltzmann solver values, and the bottom row displays the relative error percentages between the two. Remaining discrepancies in \textbf{(B)} are mainly due to non-zero numerical truncation errors in the simulated waveforms and downsampling when curve fitting. }
	\label{fig:CO2_simulation_results}
\end{figure*}

\begin{figure*}[t!]
	\centering
	\includegraphics[width=1\linewidth]{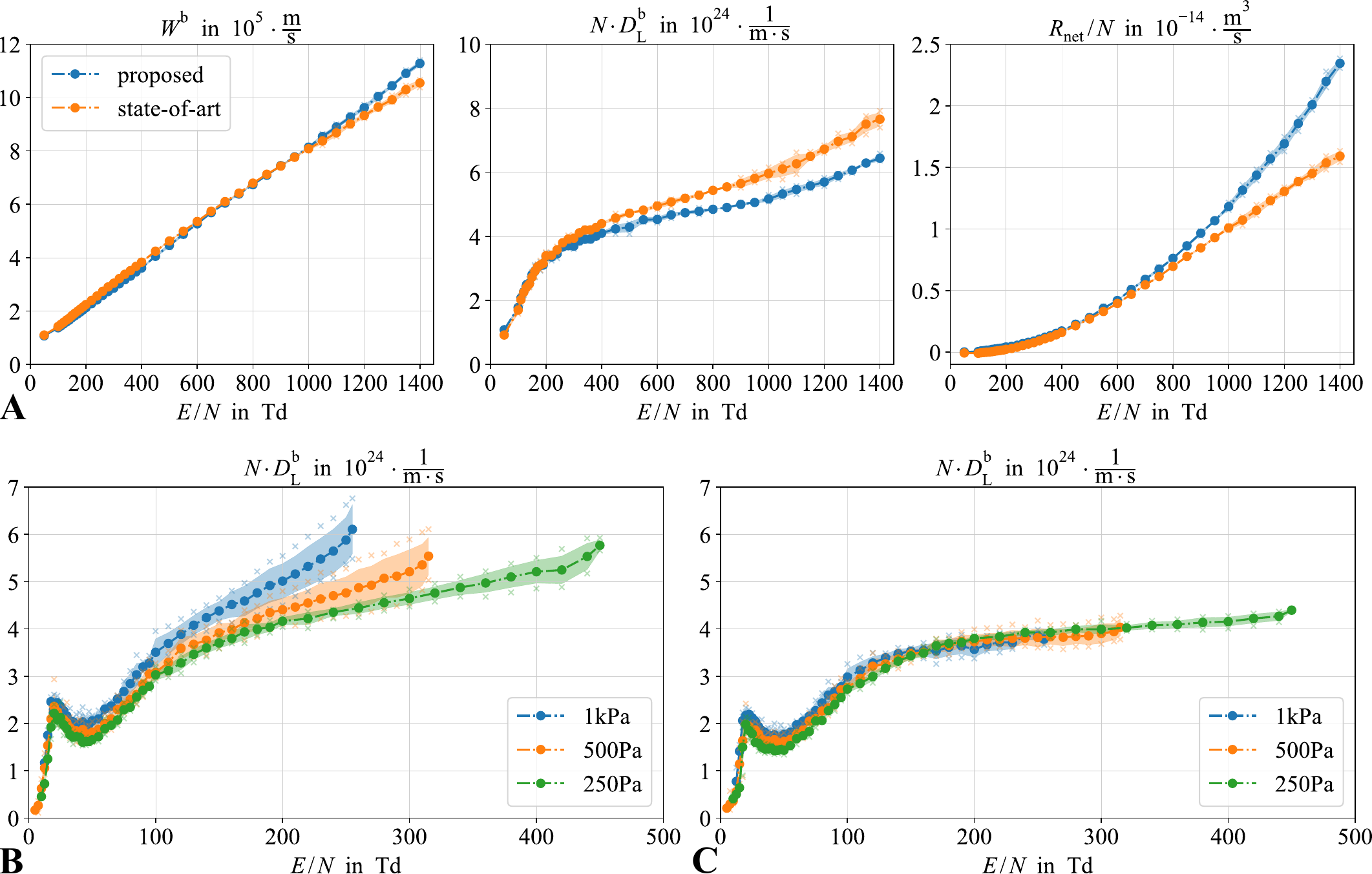}
	\caption{ \textbf{(A)} Experimental swarm parameter results of the state-of-the-art and the proposed approach for CO$\sub{2}$-N$\sub{2}$ (90\%-10\%) gas at $p = \SI{50}{Pa}$. \textbf{(B, C)} Experimental results for the reduced longitudinal diffusion coefficient $N\cdot D\sub{L}^{\mathrm{b}}$ over $E/N$ for different pressures ($p = \SI{1}{kPa},\; \SI{500}{Pa},\; \SI{250}{Pa}$) in CO$\sub{2}$-N$\sub{2}$ (90\%-10\%) gas. \textbf{(B)} Using the state-of-the-art evaluation method. \textbf{(C)} Using the proposed method. The initial swarm position was chosen as $z\sub{0} = \SI{100}{\micro\meter}$ and the pulse width around $\sigma\sub{t} = 4.5\dots\SI{5}{ns}$ (e.g., with $\SI{4.7}{ns}$) for the proposed method in \textbf{(A, C)}. Dot markers represent mean values, the shaded area shows the standard deviation, and crosses indicate minimum and maximum values. (Please note that first example swarm data, similar to the curves shown here, has already been made available by the author over LXCat \cite{CO2_LXCat}). }
	\label{fig:N2_CO2_exp}
\end{figure*}

\noindent Based on extensive evaluation of both experimental and simulated waveforms (for different gases), and by utilizing equations~\eqref{eq:back-diffusion} and \eqref{eq:convolution}, the initial electron swarm starting position $z\sub{0}$ in \textbf{Fig.~\ref{fig:electrode_arrangement}} has been determined to be around $\SI{100}{\micro\meter}$ (for the experimental Pulsed Townsend setup at ETHZ). The knowledge of an exact value for $z\sub{0}$ is not crucial and variations in the order of, e.g., $\pm50\%$ are acceptable when subsequently fitting with equation \eqref{eq:back-diffusion} (and keeping $z\sub{0}$ fixed for decreased computational complexity). This is mainly due to the extent of experimental measurement noise that distorts (the initial part of) the experimental electron current waveforms.

Generally, the physical (longitudinal) initial swarm width (denoted by $\sigma$ for the RMS width) of the initial electron swarm in \textbf{Fig.~\ref{fig:electrode_arrangement}} results from both the laser properties (e.g., laser pulse width, photon energy distribution) and the photocathode properties (e.g., work function distribution, photoemission physics). The apparent (i.e., externally observable) temporal initial swarm width (time domain) $\sigma\sub{t}$ is dominated by both the laser pulse width (in the order of a few $\SI{}{ns}$) and by the measurement bandwidth of the setup (in the order of tens of $\SI{}{MHz}$ \cite{Haefliger_thesis} and up to, e.g., $\SI{100}{MHz}$). The spatial initial swarm width (z-axis) $\sigma\sub{z}$ results from, e.g., the photon energy distribution of the laser pulse and the work function distribution of the photoelement. In practice, the impact of $\sigma\sub{t}$ is far larger than $\sigma\sub{z}$ for the typically considered measurement conditions, see section \ref{sec:improved-experimental}. 

Thus, using equations \eqref{eq:back-diffusion} and \eqref{eq:convolution}, an apparent (or total) temporal pulse width of around $\sigma\sub{t} \approx 4.5\dots\SI{5}{ns}$ is experimentally determined (and verified by simulative analysis) by curve fitting electron current waveforms at different gas pressures and varying $\sigma\sub{t}$ such that curves of $N\cdot D\sub{L}^{\mathrm{b}}$ over $E/N$ align with each other (when higher order effects such as three-body collisions are negligible, as is the case when measuring at low enough gas pressures), see section \ref{sec:improved-experimental} and \textbf{Fig.~\ref{fig:N2_CO2_exp}C}.

Lastly, both of the discussed experimental parameters in equations \eqref{eq:back-diffusion} and \eqref{eq:convolution}, $z\sub{0}$ for the initial swarm position and $\sigma\sub{t} \,(\approx\sigma)$ for the initial swarm width, are kept constant when curve fitting Pulsed Townsend measurement waveforms and extracting electron swarm transport data.

\subsection{Simulation results}

\noindent Simulated electron current waveforms are obtained for CO$\sub{2}$ gas at multiple different pressures ($p = 50,\; 500,\; \SI{5000}{Pa}$), two gap spacings ($L = 20,\; \SI{30}{mm}$) and over a wide field range ($E/N = 10\dots\SI{2000}{Td}$) by utilizing the previously mentioned fluid code, see section \ref{sec:fluid-code}. The parameters $z\sub{0}$ and $\sigma\sub{t}$ in the fluid code were chosen to closely match the experimental conditions, i.e., $z\sub{0} = \SI{100}{\micro\meter}$ and $\sigma\sub{t} = \SI{4.5}{ns}$ (and $\sigma\sub{z} = 0$ for simplicity).

\textbf{Fig.~\ref{fig:CO2_simulation_results}A} shows the extracted swarm parameters when using the existing method (equation \eqref{eq:casey-eq}). As also discussed in a separate study \cite{PT-paper-HVL} and evident from literature data \cite{Haefliger_thesis, Vass_2021, Eda_thesis, Alise_thesis}, the state-of-the-art approach fails to correctly extract $D\sub{L}^{\mathrm{b}}$ with relative errors up to $85\%$, shows large deviations for both $W^{\mathrm{b}}$ and $R\sub{net}$, and an apparent attaching region for $R\sub{net}$ at low pressures.

\textbf{Fig.~\ref{fig:CO2_simulation_results}B} presents the extracted swarm parameters when evaluating with the proposed approach (equations \eqref{eq:back-diffusion} and \eqref{eq:convolution}). The longitudinal diffusion coefficient $D\sub{L}^{\mathrm{b}}$ can in this case be accurately extracted with relative errors below around $0.15\%$ for $p = \SI{5000}{Pa}$, and $0.45\%$ for $p = \SI{500}{Pa}$, and $2.75\%$ (e.g., at $E/N = \SI{2000}{Td}$) for $p = \SI{50}{Pa}$. Similarly, the deviations in extracting $W^{\mathrm{b}}$ and $R\sub{net}$ are strongly reduced. Furthermore, the previously observed apparent attaching region in $R\sub{net}$ at low pressures is no longer visible.

\subsection{Experimental results}\label{sec:improved-experimental}

\noindent In this section, an example set of experimental measurements is further considered to verify the observed trends in the simulative analysis. In \textbf{Fig.~\ref{fig:N2_CO2_exp}A}, the swarm parameters for measurements of a CO$\sub{2}$-N$\sub{2}$ (90\%-10\%) mixture at $p = \SI{50}{Pa}$ are evaluated with both the state-of-the art approach and the proposed method. When comparing with \textbf{Fig.~\ref{fig:CO2_simulation_results}A} and \textbf{Fig.~\ref{fig:CO2_simulation_results}B}, a similar (performance) difference between the existing and proposed method can be seen in both the simulative study and in the experimental evaluation. For example, the drift velocity $W^{\mathrm{b}}$ tends to slightly overestimate up to around $E/N = \SI{900}{Td}$ and underestimate above $E/N = \SI{1000}{Td}$. Furthermore, the longitudinal diffusion coefficient $D\sub{L}^{\mathrm{b}}$ tends to overestimate in the state-of-the-art method case, as also evident from the simulation results. Lastly, the difference between the two effective ionization rates $R\sub{net}$ follows a similar trend as with the simulations, where $R\sub{net}$ tends to underestimate for the state-of-the-art approach (by up to $34\%$ at $E/N = \SI{1400}{Td}$).

In \textbf{Fig.~\ref{fig:N2_CO2_exp}BC}, the (reduced) longitudinal diffusion coefficient $N\cdot D\sub{L}^{\mathrm{b}}$ is considered for the same gas mixture as above at multiple different pressures ($p = 250,\;500,\;\SI{1000}{Pa}$). As in the simulative study, the spread of the different curves for $N\cdot D\sub{L}^{\mathrm{b}}$ is evident in the state-of-the-art method case, whereas this spread disappears for the most part for the proposed approach. Remaining slight misalignments between different pressures are mostly due to experimental measurement noise (and uncertainty), space charge effects such as electron-electron and/or electron-ion interactions at higher fields (that tend to result in a seemingly increased diffusion coefficient), the (previously) neglected spatial spread of the initial electron swarm $\sigma\sub{z}$, slight measurement loop bandwidth variations with changing electrode gap distance, and also likely due to higher order effects such as neglected higher order terms in the fluid model \eqref{eq:n_continuity}. The detailed influence of these parameters will be discussed in future work.

\section{Higher Order Effects}

\begin{figure*}[t!]
	\centering
	\includegraphics[width=1.\linewidth]{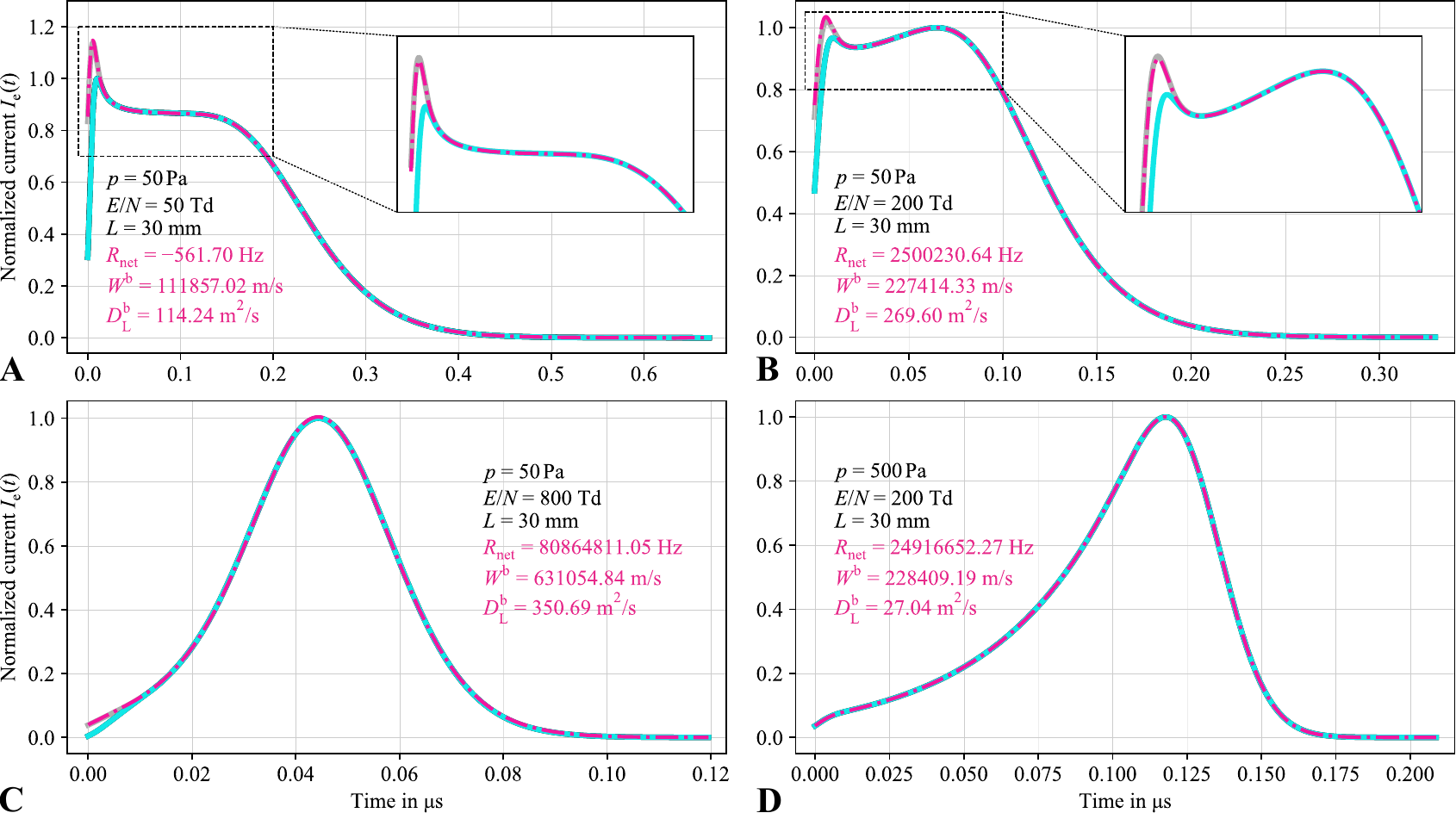}
	\caption{ A few example cases (for CO$\sub{2}$ gas) showing the possible adverse impact of the third order transport coefficient on the evaluation performance with the proposed model. The initial pulse width is chosen as $\sigma\sub{t} = \SI{4.5}{ns}$ and the initial swarm position around $z\sub{0} = \SI{100}{um}$. The blue curve {\coloredline{colorblue}} is used for fitting and represents the electron current waveform including the third order term, the pink curve {\pinkline} shows the obtained fit using the proposed method, and the underlying gray curve {\coloredline{colorgray}} depicts the current waveform that results from second order modelling (arbitrarily scaled to align with the blue waveform). \textbf{(A)} the (bulk) swarm parameters for the blue curve are chosen as: $R\sub{net} \simeq \SI{-1.109e4}{Hz}$, $W^{\mathrm{b}} \simeq \SI{1.118e5}{\meter\per\second}$, $D\sub{L}^{\mathrm{b}} \simeq \SI{1.141e2}{\meter\squared\per\second}$ and $Q\sub{L}^{\mathrm{b}} \simeq \SI{3.321e-3}{\meter\cubed\per\second}$. \textbf{(C)} the (bulk) swarm parameters for the blue curve are chosen as: $R\sub{net} \simeq \SI{8.294e7}{Hz}$, $W^{\mathrm{b}} \simeq \SI{6.272e5}{\meter\per\second}$, $D\sub{L}^{\mathrm{b}} \simeq \SI{3.736e2}{\meter\squared\per\second}$ and $Q\sub{L}^{\mathrm{b}} \simeq \SI{6.642e-2}{\meter\cubed\per\second}$. \textbf{(B)} the (bulk) swarm parameters for the blue curve are chosen as: $R\sub{net} \simeq \SI{2.491e6}{Hz}$, $W^{\mathrm{b}} \simeq \SI{2.275e5}{\meter\per\second}$, $D\sub{L}^{\mathrm{b}} \simeq \SI{2.696e2}{\meter\squared\per\second}$ and $Q\sub{L}^{\mathrm{b}} \simeq \SI{6.642e-3}{\meter\cubed\per\second}$. \textbf{(D)} as \textbf{B}, but with pressure scaling. }
	\label{fig:skewness-impact}
\end{figure*}

\noindent The impact of higher order terms in the transport equations has not been discussed so far, as these complicate the analysis and resulting evaluation procedures. However, the influence and analysis has been subject of many studies \cite{Simonovic_2020, Simonovic_2020_2, Simonovic_2022}. As such, experimentally determined values for the $3^{\text{rd}}$ order transport coefficient (longitudinal skewness $Q\sub{L}$) are rare or not available. Recently, Kawaguchi et al. \cite{Kawaguchi} presented an experimental setup and procedure to determine $Q\sub{L}$ values for gases, and provided first measurement data for N$\sub{2}$ over a wide field range of $E/N = 50\dots\SI{700}{Td}$.

In order to provide a first assessment of the adverse influence of higher order contributions such as skewness on the evaluation performance of the proposed fitting method, a few example current waveforms for CO$\sub{2}$ will be simulated using the mentioned framework in section \ref{sec:fluid-code}, and subsequently fitted using the proposed approach. \textbf{Fig.~\ref{fig:skewness-impact}} shows four different cases: three at a low pressure of $p = \SI{50}{Pa}$ with increasing field ($E/N = 50,\; 200,\;\SI{800}{Td}$), and one case at an increased pressure of $p = \SI{500}{Pa}$ ($E/N = \SI{200}{Td}$).

The low pressure, low field case shows a strong sensitivity for $R\sub{net}$ towards a non-zero $Q\sub{L}$. The higher field (and pressure) cases are less susceptible in this regard. For the considered cases, the drift velocity $W^{\mathrm{b}}$ and diffusion coefficient $D\sub{L}^{\mathrm{b}}$ are extracted fairly accurately from the 'distorted' current waveforms. However, noticeable deviations for $D\sub{L}^{\mathrm{b}}$ become visible at increasing fields (and larger $Q\sub{L}$ values).

This short analysis shows that even in the presence of a third order parameter the proposed fitting algorithm maintains good performance, and that a further detailed quantification of the influence and magnitude of higher order terms might be relevant.
A qualitative comparison between the waveforms in \textbf{Fig.~\ref{fig:example-fitting}B} and \textbf{Fig.~\ref{fig:skewness-impact}} already suggests that $Q\sub{L}^{\mathrm{b}}$ is even smaller than the conservative estimate considered here, e.g., $N^2 \cdot Q\sub{L}^{\mathrm{b}} = N^2 \cdot Q\sub{L}^{\mathrm{f}} = \SI{5e41}{\per\meter\cubed\per\second}$ for $E/N = \SI{50}{Td}$.
Future work will discuss third order and further terms in more detail and present approaches to extract these from swarm experiments (e.g., current waveforms of a Pulsed Townsend experiment).

\section{Conclusion}

\noindent In this paper, an improved evaluation approach for Pulsed Townsend experimental data is presented by providing a new analytical electron current expression and by considering and suitably implementing a finite temporal initial pulse width that accounts for both finite laser pulse width and measurement bandwidth limitations. Thus, improved swarm parameter accuracy is achieved even when modest hardware (i.e., lasers with pulse widths in the $\SI{}{ns}$ range, and common high-speed amplifiers) is used.

Specifically, simulation and experimental results show that large improvements in extracting swarm parameters are achieved, e.g., the effective ionization rate $R\sub{net}$ is improved by around $40\%$ (at high fields of, e.g., around $E/N = 1400\dots\SI{2000}{Td}$ for CO$\sub{2}$-rich gas, or equivalently under large net ionizing conditions), and measuring the longitudinal diffusion coefficient $D\sub{L}^{\mathrm{b}}$ is made practically viable.

The results and implications are not limited to CO$\sub{2}$ (or N$\sub{2}$) gas, which only serves as an example here. Furthermore, the underlying modelling principles may be applicable to other experimental configurations or measurement approaches. In particular, considering an (effective) temporal response or initial pulse width in Time-of-Flight experiments, including scanning drift tubes \cite{Korolov_2016}, could improve the determination of swarm parameters, especially $D\sub{L}^{\mathrm{b}}$. More accurate electron swarm parameter data enables more precise modelling, simulation and optimization for plasma physicists, engineers, and designers of, e.g., plasma devices, high-voltage insulation or medical equipment.

Limitations of the presented approach mainly include the neglected initial spatial pulse width $\sigma\sub{z}$, neglected higher order terms (e.g., longitudinal skewness $Q\sub{L}$), and edge effects arising from the finite electrode size. The latter become noticeable at larger transverse diffusion coefficients and smaller drift velocities, corresponding to large values of $D\sub{T}\cdot L/(W\cdot d^2)$, where $d$ represents the available lateral clearance; $D\sub{L}$ is not explicitly included in this estimate.
In practice such edge effects are typically observed at lower gas densities.
Additional limitations include space-charge effects at high fields, and potential non-hydrodynamic effects during the equilibration period of the initial electron swarm.
Future work will, thus, study the influence of higher order transport parameters (such as longitudinal skewness $Q\sub{L}^{\mathrm{b}}$ or kurtosis) on measurement results in more detail, and develop analysis techniques that consider further influences such as $\sigma\sub{z}$ and higher order effects in order to, e.g., accurately determine $Q\sub{L}$ data from swarm experiments. Improved analysis techniques for other experimental arrangements and a detailed comparison of these will also be part of future efforts. Lastly, simple and accessible methods to extract parameters beyond the one-dimensional model (e.g., the transverse diffusion coefficient $D\sub{T}$) are subject of on-going research and will be presented in future contributions.

\bibliographystyle{IEEEtran}
\bibliography{bibliography.bib}

\end{document}